\documentclass[prb,twocolumn,showpacs,superscriptaddress,floatfix,amsmath,amssymb]{revtex4}

\usepackage{graphicx} % Include figure files
\usepackage[large]{subfigure}
\usepackage{color}
\usepackage{dcolumn} % Align table columns on decimal point
\usepackage{bm} % bold math\newcommand{\N}{{\Bbb N}}   % N
\begin{document}

\title{Phase diagram of the $t$-$U$-$J_1$-$J_2$ chain at half filling}

%\date{February 15, 2008}
\date{\today}

\author{X.\ Huang}
\affiliation{Fachbereich Physik, Philipps-Universit\"at Marburg,
D-35032 Marburg, Germany}
\author{E.\ Szirmai}
\affiliation{Research Institute for Solid State Physics and Optics, H-1525
Budapest, P.O.\ Box 49, Hungary }
\author{F.\ Gebhard}
\affiliation{Fachbereich Physik, Philipps-Universit\"at Marburg,
D-35032 Marburg, Germany}
\author{J.\ S\'olyom}
\affiliation{Research Institute for Solid State Physics and Optics, H-1525
Budapest, P.O.\ Box 49, Hungary }
\author{R.M.\ Noack}
\affiliation{Fachbereich Physik, Philipps-Universit\"at Marburg,
D-35032 Marburg, Germany}

\begin{abstract}
We investigate the half-filled Hubbard chain with additional
nearest- and next-nearest-neighbor spin exchange, $J_1$ and $J_2$,
using bosonization and the density-matrix renormalization group. %RMN methods.
For $J_2=0$ we find a spin-density-wave phase for all positive values
of the Hubbard interaction $U$ and the Heisenberg exchange $J_1$.
A frustrating spin exchange $J_2$ induces a bond-order-wave
phase. For some values of $J_1$, $J_2$ and $U$, we observe
a spin-gapped metallic Luther-Emery phase.
\end{abstract}

\maketitle

\section{Introduction}
\label{sec:introduction}
The Hubbard chain is the archetype of one-dimensional strongly
correlated electron systems.
At half band-filling and for
all values of the Hubbard interaction $U$, 
it exhibits insulating spin-density-wave (SDW)
behavior, marked by a critical behavior of the spin correlations.
In a weak-coupling picture, this insulating behavior is generated by
umklapp scattering, while in strong coupling, the opening of the
Mott-Hubbard gap leads to behavior of the spin degrees of freedom
governed by an effective Heisenberg chain.
These perturbative results are reinforced by the exact Bethe-Ansatz
solution.\cite{lieb_wu_1968} Hubbard-type models
are relevant to a wide variety of one-dimensional
materials, such as polymers,\cite{Kiessbook}
cuprates,\cite{kim:137402} or TTF-TCNQ.\cite{florian-eric-holger}

One important experimental question is to what extent the spin
correlations remain critical when additional interactions are present.
It is well known that any dimerization or sufficiently large
frustration can lead to a spin gap in the Heisenberg-type spin models.
A nearest-neighbor Coulomb repulsion,
\cite{nakamura_tuv,sandvik_2004} an alternating local
potential,\cite{fabrizio_1999,manmana_2004} or a 
second-neighbor hopping\cite{fabrizio_1996,daul_2000} can lead to a
spin gap in models for
itinerant interacting  electrons. 
%{\bf RMN: more refs?}

The Hubbard model with a nearest-neighbor
antiferromagnetic exchange in two dimensions is of interest in the
context of the high-$T_c$ cuprates.
In particular, spin-liquid states~\cite{anderson}
and gossamer superconductivity~\cite{laughlin}
at and near half-filling have been proposed as necessary precursors to
high-temperature superconductivity at higher doping.
Since it is not clear whether such states are present in sufficient
strength and for sufficiently wide parameter regimes in the pure
Hubbard or $t$-$J$ models, additional interactions, including a spin
exchange, have been proposed to
be relevant.\cite{arrachea_2005}
% {\bf RMN: refs?}

In this work, we investigate the effect of two additional terms on the phase
diagram of the half-filled (average electron occupation 
$\langle n \rangle = 1$) Hubbard chain, namely
explicit antiferromagnetic exchange interactions between nearest neighbors
and between next-nearest neighbors. 
The Hamiltonian is given by
\begin{flalign}
\label{eq:Ham1}
H=-t\sum_{i,\sigma}\left(c_{i,\sigma}^\dagger c_{i,\sigma}^{\phantom{\dagger}}
 + h.c.\right) + U\sum_i n_{i,\uparrow} n_{i,\downarrow}  \nonumber \\ 
+ J_1 \sum_i {\bm S}_i {\bm S}_{i+1} + J_2 \sum_i {\bm S}_i {\bm S}_{i+2},
\end{flalign}
where $c_{i,\sigma}^\dagger$ ($c_{i,\sigma}$) creates (annihilates) an
electron with spin $\sigma$ at site $i$, $n_{i,\sigma} = c_{i,\sigma}^\dagger
c_{i,\sigma}^{\phantom{\dagger}}$, and ${\bm S}_i$ is the spin operator on
site $i$: $S_i^\alpha =
\frac12 \sum_{\sigma, \sigma'} c_{i,\sigma}^\dagger
\hat{\sigma}_{\sigma,\sigma'}^{\,\alpha} c_{i,\sigma'}^{\phantom{\dagger}}$. 
The index $\alpha=x,y,z$, and $\hat{\sigma}_{\sigma, \sigma'}^{\,\alpha}$ are
the Pauli matrices.  Here $t$ is the hopping amplitude 
% (we set $t=1$ for simplicity in the remainder of this article) 
and $U$ the strength of
the on-site Coulomb interaction.  The antiferromagnetic Heisenberg parameters
$J_1$ and $J_2$ correspond to nearest- and next-nearest-neighbor exchange,
respectively.

The unfrustrated ($J_2=0$) version of this model has previously been
investigated both analytically and numerically.
In particular, a generalized model with an anisotropic Heisenberg
coupling was investigated in Ref.\ \onlinecite{japaridze_2000} using
bosonization. 
While this work concentrated primarily on the case of ferromagnetic
exchange, isotropic antiferromagnetic exchange was included 
in a phase that is marked as `dimer long-range order',
which corresponds to a bond-order wave (BOW) in our notation; see below.
The phase diagram from bosonization of the isotropic antiferromagnetic
exchange was considered explicitly in Refs.\ \onlinecite{dai_2004} and
\onlinecite{feng_2004}, supported by numerical calculations
using the transfer-matrix renormalization group (TMRG) 
\cite{dai_2004} and exact diagonalization.\cite{feng_2004}
The %RMN alleged 
phase diagram found contains two phases: a bond charge-density-wave
phase (our BOW phase) at sufficiently small $U$ for all $J_1$, and a
SDW at larger $U$.
The critical value of $U_c$ goes to zero at small and large $J_1$
and reaches a maximum value $U_c/t \approx 0.35$ 
at intermediate $J_1$.

In this work, we reexamine the bosonization treatment of the 
$t$-$U$-$J_1$ model
in the weak-coupling regime, including the
renormalization of the coupling constants within the
mean-field approximation.
In addition, we consider the effect of the additional frustrating
exchange $J_2$, which allows us to explicitly induce the bond-order
phase and to make contact with the known phase diagram of the
frustrated Heisenberg chain at large $U$.
We also carry out high-precision ground-state density-matrix
renormalization group (DMRG) calculations, which allows us to explore
the phase diagram numerically exactly.
Both the revised bosonization and the DMRG calculations indicate that
a BOW phase is not present for $J_2=0$; the system is in a SDW phase
for {\em all\/} positive $J_1$ and $U$.
We show that a BOW phase can be induced by turning on $J_2$
positively, with the critical value required depending on $U$ and
$J_1$.
At larger values of $J_2$, we find additional phases, including a
spin-gapped metallic phase which we identify as a Luther-Emery phase.

The paper is organized as follows:
In Sec.\ \ref{sec:bosonization}, we discuss the bosonization
calculation and the resulting phase diagram.
Sec.\ \ref{sec:DMRG} contains our numerical DMRG results and
compares and contrasts the behavior obtained with that predicted by
bosonization.
In Sec.\ \ref{sec:discussion}, we discuss the overall phase diagram of
the model in terms of the results from the two methods as well as the
implications of our findings.

\section{Field theory}
\label{sec:bosonization}
% \allowdisplaybreaks
%\subsection{The Hamiltonian of the $t$-$U$-$J_1$-$J_2$ model}
We start our investigation with an analytical treatment
of our model for small couplings,
$U,J_1,J_2\ll t$. For simplicity, we % use everywhere the unit 
take $\hbar \equiv 1$ everywhere.

\subsection{Linearization of the spectrum}

In terms of fermion operators the Hamiltonian \eqref{eq:Ham1} has the form
\begin{multline}
\label{eq:Ham2}
 H=-t\sum_{i,\sigma}\left(c_{i,\sigma}^\dagger c_{i,\sigma}^{\phantom{\dagger}}
 + H.c. \right) +
 U\sum_i c_{i,\uparrow}^\dagger c^{\phantom{\dagger}}_{i,\uparrow}
 c_{i,\downarrow}^\dagger c_{i,\downarrow}^{\phantom{\dagger}} \\ + 
\sum_{\ell=1}^2
\frac{J_{\ell}}{4}\sum_i \Bigl[ 2\left(
 c_{i,\downarrow}^\dagger c_{i,\uparrow}^{\phantom{\dagger}} 
c_{i+\ell,\uparrow}^\dagger c_{i+\ell,\downarrow}^{\phantom{\dagger}} +
 c_{i,\uparrow}^\dagger c_{i,\downarrow}^{\phantom{\dagger}}
 c_{i+\ell,\downarrow}^\dagger c_{i+\ell,\uparrow}^{\phantom{\dagger}} 
\right) \\ +
 c_{i,\uparrow}^\dagger c_{i,\uparrow}^{\phantom{\dagger}}
 c_{i+\ell,\uparrow}^\dagger c_{i+\ell,\uparrow}^{\phantom{\dagger}} + 
c_{i,\downarrow}^\dagger c_{i,\downarrow}^{\phantom{\dagger}}
 c_{i+\ell,\downarrow}^\dagger c_{i+\ell,\downarrow}^{\phantom{\dagger}} \\ 
- c_{i,\uparrow}^\dagger
 c_{i,\uparrow}^{\phantom{\dagger}} c_{i+\ell,\downarrow}^\dagger 
c_{i+\ell,\downarrow}^{\phantom{\dagger}} -
 c_{i,\downarrow}^\dagger c_{i,\downarrow}^{\phantom{\dagger}} 
c_{i+\ell,\uparrow}^\dagger c_{i+\ell,\uparrow}^{\phantom{\dagger}} \Bigr] .
\end{multline}
For low temperatures and for excitations at low energies, it is enough to
consider a restricted Hilbert space which contains only states close
to the Fermi surface.  In one dimension, the Fermi surface consists
only of two points, $k=\pm k_\textrm{F}$. Around the Fermi points, the
spectrum can be linearized and one can introduce left-moving and
right-moving fermions corresponding to the states
near $-k_\textrm{F}$ and $+k_\textrm{F}$, respectively,
%\begin{flalign}
%c_{i,\sigma} \rightarrow c_{i,\sigma,+} \textrm{e}^{\textrm{i}k_\textrm{F}
%R_i} + c_{i,\sigma,-} \textrm{e}^{-\textrm{i}k_\textrm{F} R_i},
%\end{flalign}
\begin{flalign}
  c_{i+\ell,\sigma} \rightarrow  c_{i+\ell,\sigma,+}
  \textrm{e}^{\textrm{i}k_\textrm{F} (R_i+\ell a)} + c_{i+\ell,\sigma,-}
  \textrm{e}^{-\textrm{i}k_\textrm{F} (R_i+\ell a)}
\end{flalign}
for $\ell=0,1,2$.
Here $R_i$ is the coordinate vector of the site $i$ and $a$ is the
lattice constant.  For the half-filled system, $k_\textrm{F} = \pi/2a$.
Therefore, the left- and right-moving fermions have the phase factor
$\textrm{e}^{\pm \textrm{i}\ell \pi/2}$, for different
values of $\ell$. When written in terms of the chiral fermions
$c_{i+\ell,\sigma,\pm}$, each interaction  
term of Hamiltonian \eqref{eq:Ham2} splits into four new terms.
Two of them correspond to forward-scattering processes whose
couplings are denoted by $g_2$ and $g_4$ in standard
$g$-ology notation.\cite{Solyomreview}
In addition, there are two backward-scattering processes which
describe ``true'' backward scattering ($g_1$-processes) and umklapp
scattering ($g_3$-processes). 
Due to the SU(2) symmetry of the spin sector, all
processes depend only on the relative spins of the scattering
electrons. This is denoted by the subscripts $||$ and $\perp$ if the
scattering electrons have the same or opposite spins, respectively.
The relation between the $g$-ology parameters and the couplings of our
original model is
\begin{subequations}
\begin{flalign}
g_{1\perp} = & \, U - J_1/2 - 3J_2/2, \\
g_{2\perp} = & \, U + J_1/2 - 3J_2/2, \\
g_{3\perp} = & \, U + 3J_1/2 - 3J_2/2, \\
g_{4\perp} = & \, U - 3J_1/2 - 3J_2/2,
\end{flalign}
\end{subequations}
and
\begin{subequations}
\begin{flalign}
g_{1\parallel} = & \, - J_1/2 + J_2/2, \\
g_{2\parallel} = & \,  J_1/2 + J_2/2, \\
g_{3\parallel} = & \, - J_1/2 + J_2/2, \\
g_{4\parallel} = & \,  J_1/2 + J_2/2. 
\end{flalign}
\end{subequations}
In order to analyze the low-energy $g$-ology model,
we apply the bosonization method.

\subsection{Bosonization of the Hamiltonian}

First, we introduce the continuous chiral fermion fields
$\psi_{\sigma,\pm}(x)$ by making the replacement
$c_{i,\sigma,\pm}/\sqrt{a} \rightarrow \psi_{\sigma,\pm}(x)$. The
bosonization of the on-site interaction is straightforward. 
%RMN In the 
Using Abelian bosonization, %RMN technique one 
we introduce the chiral
boson phase fields $\phi_{\sigma,\pm}(x)$ via 
\begin{equation}
\psi_{\sigma,\pm}(x) = \frac{1}{\sqrt{2\pi}} F_\pm \textrm{e}^{\pm \textrm{i}
2 \phi_{\sigma,\pm}(x)},
\end{equation}
where $F_\pm$ are the so-called Klein factors which ensure the
anti-commutation relations of the fermion fields. The symmetric and
antisymmetric combination of the spin-dependent boson fields,
$\phi_{c,\pm} = \phi_{\uparrow,\pm} + \phi_{\downarrow,\pm}$ and
$\phi_{s,\pm} = \phi_{\uparrow,\pm} - \phi_{\downarrow,\pm}$,
correspond to the collective charge and spin modes, respectively. 

In order to bosonize the non-local processes, one must expand the fermion
fields with respect to the lattice constant. 
The bosonized form of the $g$-ology Hamiltonian density, up to
leading order in the expansion with respect to the lattice constant, is
\begin{flalign}
\label{eq:bosHam0}
 H^{(0)}(x) = & \, \frac{1}{2\pi} \sum_{r=\pm} \left[ v_\rho (\partial_x
 \phi_{c,r})^2 + v_\sigma (\partial_x \phi_{s,r})^2 \right] \nonumber \\ + &
 \, \frac{g_\rho}{2\pi^2}(\partial_x \phi_{c,+})(\partial_x \phi_{c,-}) -
 \frac{g_c}{2\pi^2} \cos(2\phi_c) \nonumber \\ - & \,
 \frac{g_\sigma}{2\pi^2}(\partial_x \phi_{s,+})(\partial_x \phi_{s,-}) +
 \frac{g_s}{2\pi^2} \cos(2\phi_s) \nonumber \\ - & \, \frac{g_{cs}}{2\pi^2}
 \cos(2\phi_c)\cos(2\phi_s).
\end{flalign}
Here $\phi_{c/s}=\phi_{c/s,+}+\phi_{c/s,-}$ are the total phase fields, and the
couplings are given by
\begin{subequations}
\begin{flalign}
g_\rho = & \, g_{2\perp} + g_{2\|} -g_{1\|} = U +3J_1/2-3J_2/2, \\ 
g_\sigma = & \, g_{2\perp} - g_{2\|} + g_{1\|} = U -J_1/2 - 3J_2/2, \\
g_c = & \, g_{3\perp} =  U +3J_1/2-3J_2/2, \\ 
g_s = & \, g_{1\perp}  = U -J_1/2 - 3J_2/2, \\
g_{cs}= & \, g_{3\|} = -J_1/2 + J_2/2.
\end{flalign}
\end{subequations}
The renormalized Fermi velocities are $v_\rho = 2t + (g_{4\|} +
g_{4\perp})/2\pi$ and $v_\sigma = 2t + (g_{4\|} - g_{4\perp})/2\pi$. Here and
in the following, we use the lattice constant as the unit for the 
coupling constants as well as for the Fermi velocities.

The spin-charge coupling term with coupling constant $g_{cs}$ describes
umklapp scattering processes between electrons with the same spin. This
interaction term formally occurs in the zeroth order of the expansion of the
fermion fields with respect to the lattice constant. It is clear, however,
that $g_{3\|}$ type processes can give contributions only for non-local
interactions. Moreover, this spin-charge coupling term breaks the global spin
SU(2) symmetry of the system. Therefore, in order to preserve this symmetry,
and in order to treat the non-local interactions in a consistent way, the
next-to-leading terms have to be taken into account in the expansion of the
fermion fields. To first order, among other contributions, three new
spin-charge coupling terms appear in the Hamiltonian. We find that the spin
and charge velocities are changed by the term $(- g_{1\|}/2)$, and the
symmetry-restoring non-local interaction terms are given by
\begin{flalign}
  \label{eq:Ham_nextorder}
H^{(1)}(x) = & \, \frac{g_{c \sigma}}{2\pi^2} (\partial_x \phi_{s,+})
  (\partial_x \phi_{s,-}) \cos(2\phi_c) \nonumber \\ - & \, \frac{g_{\rho
  s}}{2\pi^2} (\partial_x \phi_{c,+}) (\partial_x \phi_{c,-}) \cos(2\phi_s)
  \nonumber \\ + & \, \frac{g_{\rho \sigma}}{2\pi^2} (\partial_x \phi_{c,+})
  (\partial_x \phi_{c,-})(\partial_x \phi_{s,+}) (\partial_x \phi_{s,-}).
\end{flalign}
The first two terms correspond to backward and umklapp scattering,
respectively, between electrons with opposite spins, and the third term
describes backward-scattering processes between electrons with equal
spins. 
Initially, all these couplings are equal to $g_{cs}$,
\begin{equation}
g_{\rho s}=g_{c\sigma}=g_{\rho\sigma}=g_{cs}= -J_1/2 + J_2/2 .
\end{equation}
The SU(2) symmetry of the spin sector assures %RMN the relations that
$g_s=g_\sigma$, $g_{cs}=g_{c\sigma}$, and $g_{\rho s} =g_{\rho
\sigma}$. Therefore, there are five independent couplings which we choose to
be $g_\rho$, $g_c$, $g_s$, $g_{cs}$, and $g_{\rho s}$.  We note that the
renormalization of the Fermi velocities, which is a secondary effect, will 
not be taken into account in the following.

\subsection{Renormalization group analysis 
for fluctuating charge and spin fields}

The Hamiltonian $H=H^{(0)}+H^{(1)}$ cannot be solved exactly. However,
a renormalization group (RG) analysis permits the investigation of
the relative importance of the various couplings. 
In the RG procedure, the couplings are considered to be a function of some
scaling parameter $y$, e.g., the logarithm of the effective bandwidth.
As the scaling parameter is taken to infinity, the flow of the couplings
shows which of them are important and which can be ignored,
depending on whether or not they tend to zero, to a finite value, 
or to infinity. For example, when all couplings 
but the forward scattering terms tend to zero,
the Hamiltonian~$H$ describes a Luttinger liquid with freely propagating
charge and spin degrees of freedom.

The one-loop RG equations for our five dimensionless running coupling
constants $\tilde{g}_x(y) \equiv g_x(y)/4\pi t$
read~\cite{Tsuchiizu_tuv,dai_2004}
\begin{subequations}
\label{eq:RG_cs}
\begin{flalign}
\frac{\textrm{d}\tilde{g}_\rho(y)}{\textrm{d} y} = & \, 2 \tilde{g}_c^2 +
\tilde{g}_{cs}^2 + \tilde{g}_s \tilde{g}_{\rho s} , \\
\frac{\textrm{d}\tilde{g}_c(y)}{\textrm{d} y} = & \, 2 \tilde{g}_\rho
\tilde{g}_c - \tilde{g}_s \tilde{g}_{cs} - \tilde{g}_{cs} \tilde{g}_{\rho s} ,
\\ \frac{\textrm{d}\tilde{g}_s(y)}{\textrm{d} y} = & \, - 2 \tilde{g}_s^2 -
\tilde{g}_c \tilde{g}_{cs} - \tilde{g}_{cs}^2, \\
\frac{\textrm{d}\tilde{g}_{cs}(y)}{\textrm{d} y} = & \, - 2 \tilde{g}_{cs} + 2
\tilde{g}_\rho \tilde{g}_{cs} - 4 \tilde{g}_s \tilde{g}_{cs} - 2 \tilde{g}_c
\tilde{g}_s \nonumber \\ & \, - 2 \tilde{g}_c \tilde{g}_{\rho s} - 4
\tilde{g}_{cs}\tilde{g}_{\rho s} , \\ \frac{\textrm{d}\tilde{g}_{\rho
s}(y)}{\textrm{d} y} = & \, -2 \tilde{g}_{\rho s} + 2 \tilde{g}_\rho
\tilde{g}_s - 4 \tilde{g}_c \tilde{g}_{cs} - 4 \tilde{g}_{cs}^2 \nonumber \\ &
\, - 4 \tilde{g}_s \tilde{g}_{\rho s} \, ,
\end{flalign}
\end{subequations}
with initial values $\tilde{g}_x(y=0)=g_x/4\pi t$. From these equations,
it follows that there is only a single line
of weak-coupling fixed points, namely 
$\overline{g}_{c}=\overline{g}_{s}=\overline{g}_{cs}=\overline{g}_{\rho s}=0$. 
In order to show this, we note that we have
started our analysis assuming that there is neither
a charge gap nor a spin gap. This implies that a weak-coupling fixed point
corresponds to $\overline{g}_{c}=\overline{g}_{s}=0$. 
%The equations
Equations~\eqref{eq:RG_cs} immediately imply that 
$\overline{g}_{cs}= \overline{g}_{\rho s}=0$ also, and that only
$\overline{g}_{\rho}$ remains undetermined.

A linear stability analysis of the fixed-point line shows that
it is stable against
small perturbations $g_{cs}$ and $g_{\rho s}$, that it is
marginally stable against small perturbations 
$g_s$ and $g_\rho$, and that its stability with respect to
perturbations $g_c$ depends on the sign of the fixed-point value
$\overline{g}_{\rho}$ (stable for $\overline{g}_{\rho}<0$,
unstable for $\overline{g}_{\rho}>0$).
Therefore, in order to determine the weak-coupling regime, 
it is convenient and sufficient 
to consider the RG equations without the spin-charge
coupling terms, i.e., we may consider the RG equations
for $\tilde{g}_{cs}=\tilde{g}_{\rho s}=0$.
We thus arrive at 
\begin{subequations}
\begin{flalign}
\frac{\textrm{d}\tilde{g}_\rho(y)}{\textrm{d} y} = & \, 2 \tilde{g}_c^2 ,
\label{gc-flow1}
\\ \frac{\textrm{d}\tilde{g}_c(y)}{\textrm{d} y} = & \, 2 \tilde{g}_\rho 
\tilde{g}_c ,
\\ 
\frac{\textrm{d}\tilde{g}_s(y)}{\textrm{d} y} = & \, - 2 \tilde{g}_s^2
\end{flalign}
\end{subequations}
in the vicinity of the weak-coupling fixed-point line. 

This simpler problem
is readily analyzed.
The trajectory for the spin coupling $\tilde{g}_s(y)$ flows
to infinity if $g_s < 0$. In this case, a gap opens in the spin spectrum.
If $g_s > 0$, this coupling is marginally irrelevant, i.e.,
the spin mode remains soft.
In the charge sector, $g_{\rho} = g_c$ initially,
and this relation remains valid under the RG flow. Therefore, 
it is sufficient to %RMN address 
consider Eq.~(\ref{gc-flow1}). It is seen that for
$g_c > 0$ the charge mode becomes gapped because $\tilde{g}_c(y)$
flows to infinity, otherwise the charge excitations remain gapless.

The simplified equations show that a 
fully gapless Luttinger-liquid phase, $\overline{g}_c=\overline{g}_s=0$,
is not possible for our model. 
The initial couplings would have to fulfill $g_c<0$ and $g_s>0$
which requires $J_2 > 2U/3 + J_1$ for $g_c<0$ and 
$J_2 < (2U - J_1)/3$ for $g_s>0$. These two conditions 
cannot be fulfilled simultaneously with positive bare couplings 
$U$, $J_1$, and $J_2$.
Consequently, we must redo our RG analysis under the assumption that
at least one of the two modes is gapped.

\subsection{Renormalization group analysis 
for gapped charge and/or spin modes}

When one of the fields is gapped, the spin-charge coupling processes
become relevant.\cite{Gogolinbook,Tsuchiizu_tuv}
Their contribution will be
considered on the mean-field level. 
In this picture, %RMNcase 
the gapped field is locked to a value which 
%in order to 
optimizes the
interaction energy.

When there is a gap in the charge sector,
the charge field $\phi_c$ is locked at $\overline{\phi}_c=0\, {\rm mod}\, \pi$ 
because the initial value of
the coupling $g_c$ is positive. Neglecting the fluctuations of the
field $\phi_c$ in Hamiltonian~\eqref{eq:Ham_nextorder}, the
terms proportional to $g_{\rho s}$ and $g_{\rho \sigma}$ do not contribute,
and $\cos(2\phi_c)$ can be replaced by its 
weak-coupling mean-field value, $\overline{\cos(2\phi_c)}= 1$.
Due to this substitution,
the interaction terms proportional to $g_{cs}$ and $g_{c\sigma}$ become 
marginal because their scaling dimensions reduce to $\overline{x}_{cs}=
\overline{x}_{c\sigma}=2$.
On the mean-field level, the spin-coupling term proportional to
$g_{cs}$ is of the same form
as the interaction term proportional to $g_s$  in $H^{(0)}$. 
Therefore, the spin field $\phi_s$ fluctuates in the modified potential
$g_s^* \cos(2\phi_s)$
with the new coupling $g_s^*$,
\begin{equation}
g_s^* = g_s - g_{cs} = U - 2J_2 .
\end{equation}
%RMN Along the same line of reasoning,
Analogously,
the interaction term
proportional to $g_{c\sigma}$ in $H^{(1)}$ combines with
the interaction term proportional to $g_{\sigma}$ in $H^{(0)}$
to produce the new coupling $g_{\sigma}^*$, with
\begin{equation}
g_{\sigma}^* = g_{\sigma} - g_{c \sigma} = U -2J_2 .
\end{equation}
This equation shows that the SU(2) spin symmetry
is preserved on the mean-field level.

In the presence of a charge gap and the SU(2) spin symmetry,
we only have to analyze a single equation for $\tilde{g}_s$
instead of the five RG equations~\eqref{eq:RG_cs}, namely
\begin{equation}
\frac{\textrm{d}\tilde{g}_s(y)}{\textrm{d} y} = - 2 \tilde{g}_s^2 \, ,
\end{equation}
with the initial value $\tilde{g}_s(y=0)=g_s^*/4\pi t$.
It is readily seen that the spin
mode becomes gapped if $g_s^*<0$, i.e., 
$J_2 > U/2$, independently of the value of the nearest-neighbor 
interaction $J_1$. 

\begin{figure}[ht]
\includegraphics[scale=0.35]{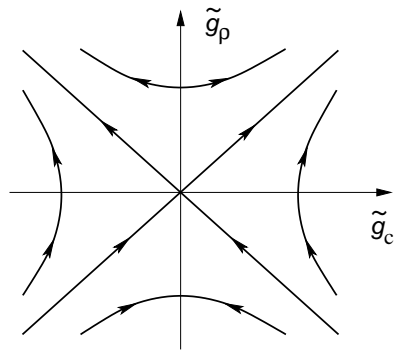}
\caption{Scaling curves for the charge-coupling parameters 
$\tilde{g}_c$ and $\tilde{g}_{\rho}$ in the presence of a spin gap.}
\label{fig:scaling-curves-1}
\end{figure}

When there is a gap in the spin sector,
the spin field $\phi_s$ is locked at $\overline{\phi}_s=0\, {\rm mod}\, \pi$ 
because the initial value of the coupling $g_s$ is negative. 
Neglecting the fluctuations of the
field $\phi_s$ in the Hamiltonian~\eqref{eq:Ham_nextorder}, the
terms proportional to $g_{\rho \sigma}$ and $g_{c \sigma}$ do not contribute 
and $\cos(2\phi_s)$ can be substituted by its 
weak-coupling mean-field value, $\overline{\cos(2\phi_s)}= 1$.
Due to this substitution,
the interaction terms proportional to $g_{cs}$ and $g_{\rho s}$ become 
marginal because their scaling dimensions reduce to $\overline{x}_{cs}=
\overline{x}_{\rho s}=2$.
On the mean-field level, the charge-coupling term proportional to
$g_{cs}$ is of the same form
as the interaction term proportional to $g_c$  in $H^{(0)}$. 
Therefore, the charge field $\phi_c$ fluctuates in the modified potential
$g_c^* \cos(2\phi_c)$ with the new coupling $g_c^*$,
\begin{equation}
g_c^* =  g_c + g_{cs} = U + J_1-J_2 .
\end{equation}
%RMN Along the same line of
Using similar
reasoning, the new coupling $g_{\rho}^*$
becomes
\begin{equation}
g_{\rho}^* = g_{\rho} - g_{\rho s} = U + 2J_1-2J_2 .
\end{equation}
Note that these new initial couplings are {\sl not\/} equal, so we
must analyze the two-dimensional scaling curves defined by 
the equations
\begin{subequations}
\begin{flalign}
\frac{\textrm{d}\tilde{g}_\rho(y)}{\textrm{d} y} = & \, 2 \tilde{g}_c^2 ,
% \label{gc-flow1}
\\ \frac{\textrm{d}\tilde{g}_c(y)}{\textrm{d} y} = & \, 2
\tilde{g}_\rho
\tilde{g}_c  \, ,
\end{flalign}
\end{subequations}
given the initial values $\tilde{g}_c(y=0)=g_c^*/4\pi t$
and $\tilde{g}_{\rho}(y=0)=g_{\rho}^*/4\pi t$.
The flow diagram is shown in Fig.~\ref{fig:scaling-curves-1}. 

The conditions for a gapped charge mode
are either $g_\rho^* > 0$ or ($g_\rho^* < 0$ and $|g_c^*| > |g_\rho^*|$). 
This leads to the result that a gapped charge mode exists
if $J_2 < 2U/3 + J_1$. 

\subsection{Phase diagram}

In general, we find three regions where either the charge gap or the
spin gap or both are finite. It is interesting to analyze the 
dominant correlations in the various gapped phases. 
The order parameters for density waves 
in the charge (CDW), spin (SDW), bond-charge (BCDW), and
bond-spin (BSDW) require the calculation of correlation functions
using the operators
\begin{subequations}
\begin{flalign}
\mathcal{O}_{i,\textrm{CDW}} & = (-1)^i (n_{i,\uparrow} + n_{i,\downarrow}), \\
\mathcal{O}_{i,\textrm{SDW}} & = (-1)^i (n_{i,\uparrow} - n_{i,\downarrow}), \\
\mathcal{O}_{i,\textrm{BCDW}} & = (-1)^i (c_{i,\uparrow}^\dagger 
c_{i+1,\uparrow} +
c_{i,\downarrow}^\dagger c_{i+1,\downarrow} + h.c.) , \\
\mathcal{O}_{i,\textrm{BSDW}} & = (-1)^i (c_{i,\uparrow}^\dagger 
c_{i+1,\uparrow} -
c_{i,\downarrow}^\dagger c_{i+1,\downarrow} + h.c.) \, ,
\end{flalign}
\end{subequations}
written in terms of the lattice fermions. 
These order parameters become 
\begin{subequations}
\label{eq:orderparameter}
\begin{flalign}
\mathcal{O}_\textrm{CDW}(x) & \propto \sin \phi_c (x) \cos \phi_s (x), \\
\mathcal{O}_\textrm{SDW}(x) & \propto \cos \phi_c (x) \sin \phi_s (x), \\
\mathcal{O}_\textrm{BCDW}(x) & \propto \cos \phi_c (x) \cos \phi_s (x), \\
\mathcal{O}_\textrm{BSDW}(x) & \propto \sin \phi_c (x) \sin \phi_s (x)
\end{flalign}
\end{subequations}
in bosonized form.
When the charge mode is gapped, the field $\phi_c$ is locked at 
$\overline{\phi}_c=0\, {\rm mod}\, \pi$.
When the spin mode is gapped, the field $\phi_s$ is locked
at $\overline{\phi}_s=0\, {\rm mod}\, \pi$.
Therefore, in the regime where both of the fields are
gapped, we find that the BCDW order parameter is maximal.
Therefore, the model describes a phase with
bond ordering (BOW) for $\Delta_c\neq 0$ and $\Delta_s\neq 0$.

When only the charge mode is gapped, the spin field is a free field.
However, upon increasing the scaling
parameter ($y$) of the renormalization group procedure, the 
initially negative spin coupling
grows and tends to zero, and the spin field oscillates
around $\pi/2$ (mod $\pi$). Therefore, for small couplings, 
the dominating ordering is SDW for
$\Delta_c\neq 0$ and $\Delta_s= 0$. Note that the SU(2) spin symmetry
is not spontaneously broken, i.e., the spin correlations
are critical without true long-range order. 

\begin{figure}[ht]
\includegraphics[scale=0.8]{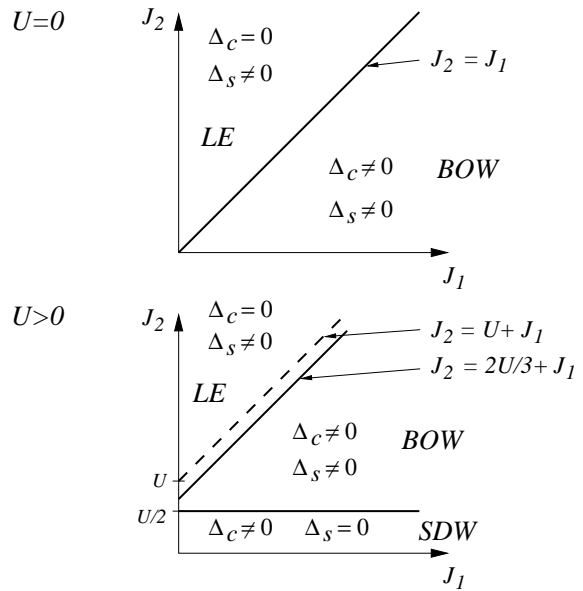}
\caption{Field-theory prediction 
for the half-filled $t$-$U$-$J_1$-$J_2$ model.
The solid lines give the phase boundaries between the fully gapped 
regime (bond-order wave, BOW) and the
semi-gapped regimes (spin-density wave, SDW; Luther-Emery, LE).
The dashed line shows the border between dominantly
charge-density-wave and bond-order-wave correlations in the
Luther-Emery phase.}
\label{fig:phases-RG}
\end{figure}

Similarly, when the spin mode is gapped and the charge mode is gapless,
there is no true long-range charge order. Therefore, we call this
phase the Luther-Emery (LE) phase.
The charge coupling $g_c$ tends to zero, either from positive values
or from negative values. Depending on the sign of the charge coupling, 
$\phi_c$ fluctuates around $\pi/2$ or around zero.
Correspondingly, the dominating correlations are either CDW or BCDW
for $\Delta_c= 0$ and $\Delta_s\neq 0$.
The line which separates the dominant BCDW critical correlation 
and the dominant CDW correlations in the LE phase is 
indicated in Fig.~\ref{fig:phases-RG} by a dashed line.

The resulting phase diagram of the $t$-$U$-$J_1$-$J_2$ 
model at weak coupling is shown in Fig.~\ref{fig:phases-RG}.
For $U=0$, the spin gap is always finite for $J_2>0$.
For $J_2<J_1$, the charge gap is also finite, and the ground state
is characterized by a bond-order wave.
The charge gap closes at $J_2=J_1$ and the system goes into
a LE phase with no long-range charge or spin ordering
but critical charge-density-wave correlations.

For $U>0$, $J_1>0$, and $J_2<U/2$, the ground state is analogous to 
the spin-density-wave (SDW) phase of the one-dimensional
Hubbard model, i.e., the charge gap is finite, the spin gap is zero,
and the spin correlations are critical.
For $2U/3+J_1>J_2>U/2$, both the spin gap and the charge gap are finite.
The ground state is a BOW with long-range order
in the bond-charge-density-wave correlations.
For $J_2>2U/3+J_1$, the charge gap closes and the system goes over
to the LE phase with a finite spin gap but no 
charge long-range order. For $2U/3+J_1<J_2<U+J_1$, the
bond-charge-density-wave fluctuations dominate, whereas, 
for $J_2>U+J_1$, the fluctuations
in the charge-density-wave order parameter are maximal.

\begin{figure}[ht]
\includegraphics[scale=0.8]{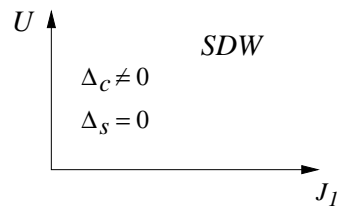}
\caption{Field-theory prediction 
for the half-filled $t$-$U$-$J_1$ model.
For all $J_1>0$, the ground state is a spin-density-wave (SDW) phase
with a finite charge gap, zero spin gap and critical spin correlations.}
\label{fig:killDai}
\end{figure}

In order to make contact with earlier work,
we display  the phase diagram of the $t$-$U$-$J_1$
model separately in Fig.~\ref{fig:killDai}. 
In contrast to previous results,\cite{japaridze_2000,dai_2004,feng_2004}
we do not find any signature of a BOW %bond-order-wave 
phase. 
For all $J_1>0$, the ground state is SDW, just
as is the ground state of the half-filled Hubbard model for $U>0$.
This result is corroborated by our numerical DMRG data, which we 
present in the next section.

\section{Numerical Results}
\label{sec:DMRG}
In order to explore the phase diagram of the Hamiltonian~(\ref{eq:Ham1}) 
and to test the predictions of bosonization, we carry out extensive,
high-precision, ground-state DMRG 
calculations.\cite{dmrg_prl_prb,dmrgbook,uli_rmp}
Relatively high sensitivity is required to resolve the phases,
especially in the weak-coupling regimes in which one would expect
bosonization to be valid.
In order to differentiate the possible phases,
we calculate the spin
gap $\Delta_s$, the charge gap $\Delta_c$, and the bond-order-wave 
parameter $\langle B \rangle$ of the one-dimensional 
$t$-$U$-$J_1$-$J_2$ model on 
lattices with open boundary conditions and up to $L=256$ sites.
The weight of the discarded density-matrix eigenstates is held below a
maximum of $10^{-9}$.

For finite systems, the spin gap $\Delta_s(L)$ is defined as
\begin{equation}
\Delta_s(L) = E_{0}(L,N,S = 1) - E_{0}(L,N,S = 0). 
\end{equation}
Accordingly, the
charge gap $\Delta_c(L)$ is determined using
\begin{eqnarray}
\Delta_c(L) &=& [E_{0}(L,N+2,S =1) + E_{0}(L,N-2,S = 0) \nonumber \\ 
            &&- 2E_{0}(L,N,S = 0)]/2,
\end{eqnarray}
where $E_{0}(L,N,S)$ is the 
ground-state energy for an $L$-site system with $N$ electrons and
total spin~$S$.
We extrapolate using second-order polynomials in $1/L$
to determine the spin gap $\Delta_s$ and the
charge gap $\Delta_c$ in the thermodynamic limit,
\begin{eqnarray}
\Delta_s(L) & = \Delta^\infty_s + A_s/L + B_s/L^2,\nonumber \\ 
\Delta_c(L) & = \Delta^\infty_c + A_c/L + B_c/L^2,\label{eq:Gap_extrapol}
\end{eqnarray}
where $\Delta^\infty_{c,s}$, $A_{c,s}$, and $B_{c,s}$ are fitting parameters.
The staggered bond order parameter is defined as
\begin{equation}
\langle B\rangle(L)=
 \frac{1}{2(L-1)}\sum_{i=1}^{L}\sum_{\sigma}
(-1)^{i+1}\langle c_{i\sigma}^{\dag}c_{i+1,\sigma} + h.c \rangle.
\end{equation}
The bond order parameter $\langle B \rangle$ is extrapolated using
finite-size corrections of the form $1/L^{\gamma}$, without considering 
higher corrections,
\begin{equation}
\langle B \rangle(L) = \langle B^\infty \rangle + A_{B}/L^{\gamma} \, ,
\label{eq:BOW_extrapol}
\end{equation}
where $\langle B^\infty \rangle$, $A_B$, and $\gamma$ are fitting parameters.
We find that adding higher-order terms, which increases the number of
fit parameters, tends to make the fits less stable.

In the following, we first treat the $t$-$U$-$J_1$ model, i.e.,
$J_2=0$ in Hamiltonian~(\ref{eq:Ham1}), then study finite positive
$J_2$, first with $U=0$, then with nonzero $U$. For simplicity, in the
remainder of this article the energy scale is set by taking $t=1$, and
so $U$, $J_1$, and $J_2$ are dimensionless quantities.

\subsection{Results for $J_2=0$} 

For the unfrustrated case ($J_2=0$), our bosonization procedure
of Sec.~\ref{sec:bosonization} predicts 
a SDW phase with a finite charge gap and critical gapless spin excitations,
$\Delta_c>0$ and $\Delta_s=0$.
In the SDW phase, the bond order parameter vanishes.

\begin{figure}
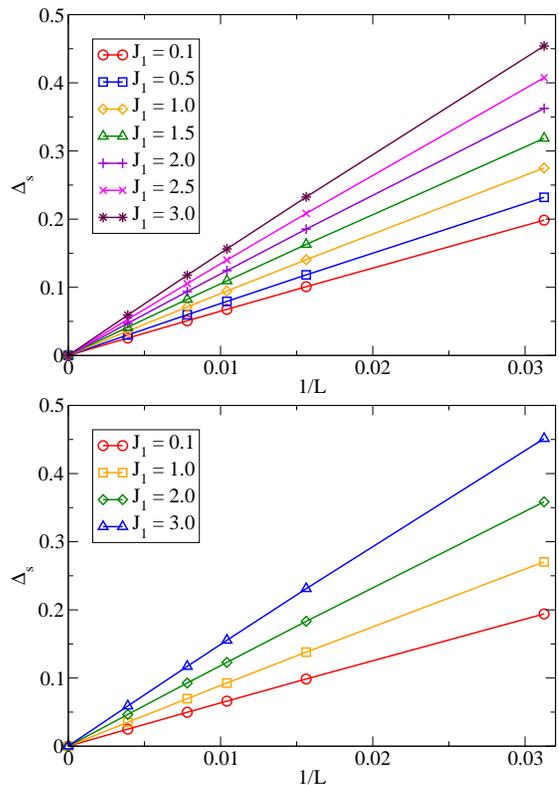

\includegraphics[scale=0.3,clip=true]{fig4a.eps}
\includegraphics[scale=0.3,clip=true]{fig4b.eps}\hfill
\caption{(Color online) Finite-size extrapolation of the spin gap as a function of
$1/L$ for the $t$-$U$-$J_1$ model at (a) $U = 0$ and (b) $U = 0.1$. }
\label{fig:spin_gap_L_U0_J20}
\end{figure}

The finite-size extrapolation of the spin gap, plotted as a function
of $1/L$ for $U=0$ and $U=0.1$ 
is shown in Fig.~\ref{fig:spin_gap_L_U0_J20}.
As can be clearly seen, the scaling behavior is predominantly linear
in $1/L$, and the $1/L\to 0$ extrapolated value, $\Delta^\infty_s$,
is zero on the scale of the plot for all values of $J_1$ for both
values of $U$.
A fit of the data with a second-order polynomial in $1/L$, as
discussed above, yields a value of
$\Delta_s$ that is less than $2\times 10^{-4}$ in all cases.
This puts a rather stringent constraint on bond ordering in this case;
the spin excitations are gapless to a very high numerical accuracy.

\begin{figure}
\includegraphics[scale=0.35,clip=true]{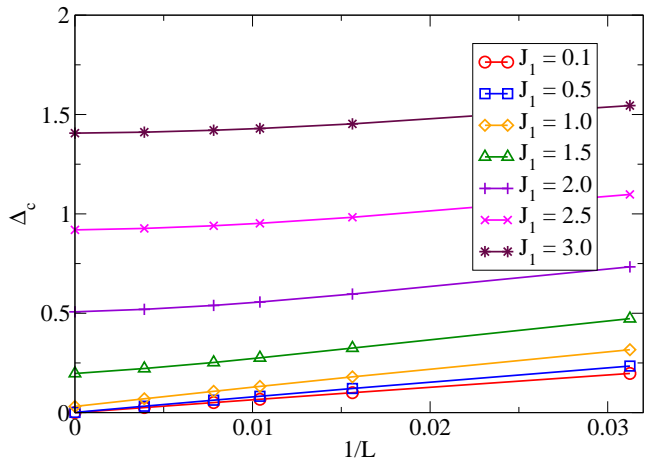}
\caption{(Color online) Finite-size extrapolation of the charge gap as a function of
$1/L$ for the $t$-$U$-$J_1$ model at $U = 0$.} 
\label{fig:charge_gap_L_U0_J20}
\end{figure}

The system-size behavior of the charge gap is displayed in 
Fig.~\ref{fig:charge_gap_L_U0_J20}.
As can be seen, the $1/L\to 0$ extrapolated value,
$\Delta^\infty_c$, is nonzero in
general, with the scaling going from being predominantly linear in
$1/L$ (with a small negative $(1/L)^2$ term) when
$\Delta_c$ is small, to having a 
substantial positive $(1/L)^2$ term when $\Delta_c$ is
significantly different from zero.
Such finite-size behavior is typical for gaps in one-dimensional
systems with open boundary conditions.

\begin{figure}
\includegraphics[scale=0.35,clip=true]{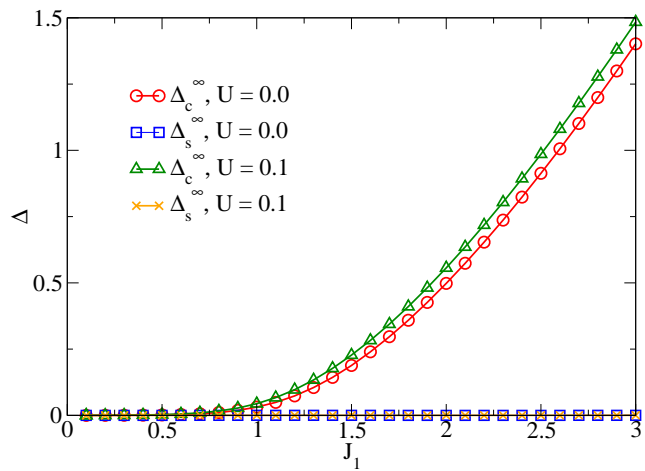}
\caption{(Color online) Extrapolated spin and charge gaps for the
$t$-$U$-$J_1$ model at $U = 0$ and 0.1 as functions of $J_1$.}
\label{fig:gaps_U0}
\end{figure}

The behavior of the extrapolated gaps as a function of $J_1$ is shown 
in Fig.~\ref{fig:gaps_U0}.
As discussed above, the spin gap is numerically indistinguishable from
zero for all values of $J_1$ for both $U=0$ and $U=0.1$.
The extrapolated charge gap is small on the scale of the plot for 
$J_1 \lesssim 0.8$, and then increases, crossing over to a linear
increase for larger values of $J_1$.
From bosonization, we would expect an exponential opening of the gap
with $J_1$, similar to the exponential opening of the charge gap with
$U$ in the $J_1=0$ case.\cite{ovchinikov_xxxx}
The $J_1$-dependence of $\Delta^\infty_c$ in 
Fig.~\ref{fig:gaps_U0} is qualitatively consistent with such a
behavior.
We have not carried out an explicit fit because the detailed form of
the exponential opening is not known from bosonization; to determine
the specifics of a general exponential form via fitting to finite-size
extrapolated data is difficult.

\begin{figure}[ht]
\includegraphics[scale=0.35,clip=true]{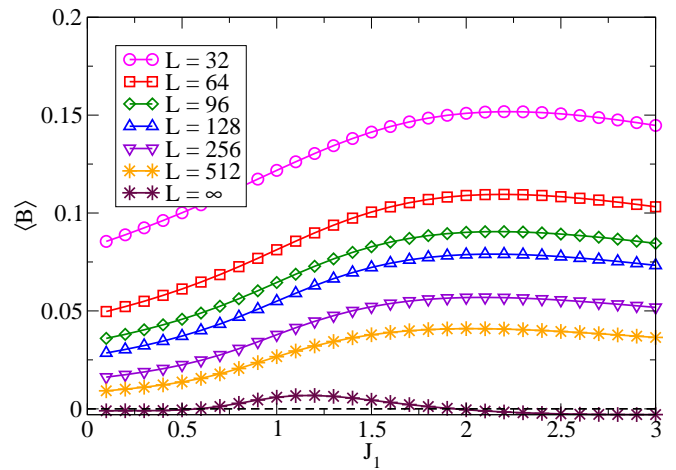}
\caption{(Color online) Bond-order parameter $\langle B\rangle(L)$
for $L = 32, 64, 96, 128, 256, 512$ and extrapolated bond order parameter 
$\langle B^\infty \rangle$ as a function of $J_1$ for 
the $t$-$U$-$J_1$ model at $U = 0$. } 
\label{fig:BOW_U0}
\end{figure}

We now turn to the BOW order parameter, displayed as a function of
$J_1$ for various system sizes and $L=\infty$ in 
Fig.~\ref{fig:BOW_U0}.
At each system size, $\langle B \rangle(L)$ has an appreciable positive finite
value which varies significantly as a function of $J_1$.
The $L\to\infty$ extrapolated value $\langle B^\infty \rangle$
is small, but still shows some variation with $J_1$.
Note, however, that the extrapolated value is negative at small and
large $J_1$ and is positive only for intermediate $J_1$.
Taking the largest negative value 
($\langle B^\infty\rangle\approx -0.003$) as a rough
estimate of the extrapolation error, the largest positive value, 
$\langle B^\infty\rangle\approx 0.007$, is not distinguishable from
zero to within our accuracy.
% This underlines the uncertainty in carrying out extrapolations using
% Eq.\ (\ref{eq:BOW_extrapol}); an example of such an extrapolations is
% shown in Fig.~\ref{fig:BOW_L_U0_J1_J2}.
% Here, the fitting yields exponents $\gamma \approx 0.47,\ldots , 0.77$ and 
% $\langle B^\infty\rangle$ is sensitive to the details of the fit.
% As 
Moreover the fit to Eq.\ (\ref{eq:BOW_extrapol}) yields an exponent
$\gamma$ which varies between $0.47$ and $0.77$. All this underlines
the uncertainty in carrying out extrapolations using this analytic form
and the sensitivity of $\langle B^\infty\rangle$ to the details of the fit. 
On the other hand, as
discussed above, $\Delta^\infty_s$ vanishes to a
high accuracy for all $J_1$, precluding a BOW phase.  
Thus, 
within the
numerical methods applied here, the spin gap seems to be a
significantly more sensitive probe for the existence of a bond order
wave phase than the bond order parameter $\langle B\rangle$ itself.

Our DMRG calculations for $J_2=0$ are thus in agreement with the
predictions of the bosonization calculations of 
Sec.~\ref{sec:bosonization}; see Fig.~\ref{fig:killDai}: 
the ground-state phase is a SDW with gapless spin
excitations for all positive $U$ and $J_1$.
While we have treated explicitly only two values of the interaction strength,
$U=0$ and $U=0.1$, we have chosen these values in
accordance with the phase diagrams of Refs.~\onlinecite{dai_2004} 
and~\onlinecite{feng_2004} which predict the appearance of a bond order wave
phase only for $U\lesssim 0.35$.
At larger values of $U$, the behavior should be that of the ordinary
half-filled Hubbard chain and one would not expect a BOW phase to
occur.

\begin{figure}
\includegraphics[scale=0.35,clip=true]{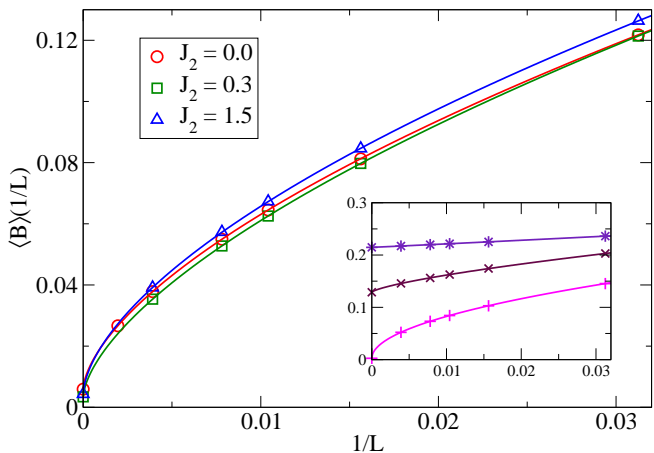}
\caption{(Color online) Finite-size scaling analysis for $\langle B \rangle(L)$ for
different $J_2$ when $U=0$ and $J_1 = 1$. 
The inset shows the finite-size scaling analysis 
for $J_2 = 2.0$, 2.5, and 3.0, from bottom to top.}
\label{fig:BOW_L_U0_J1_J2}
\end{figure}

\subsection{Results for $U=0$ and nonzero $J_2$ }
\begin{figure}
\includegraphics[scale=0.35,clip=true]{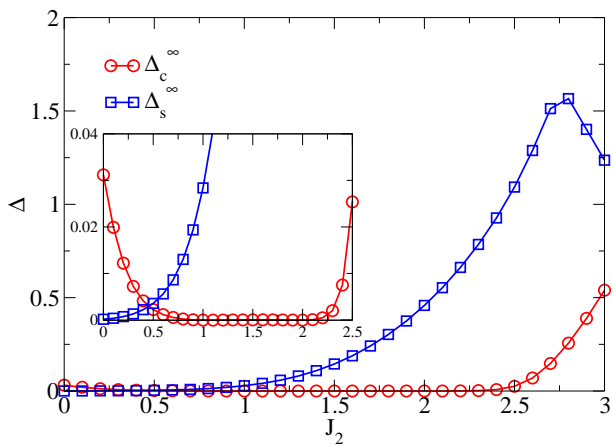}
\caption{(Color online) Extrapolated spin gap and charge gap as
functions of $J_2$ for $U=0$, $J_1=1$. The inset displays 
the same data for $J_2\leq 2.5$ on an enlarged scale.}
\label{fig:gaps_U0_J1_J2}
\end{figure}

We now include the explicit frustration $J_2$ while setting the
on-site Coulomb interaction to zero.
Fig.\ \ref{fig:gaps_U0_J1_J2} shows the system-size extrapolated spin
and charge gaps, $\Delta^\infty_s$ and $\Delta^\infty_c$, as 
functions of $J_2$ at $U=0$ and $J_1=1$.
(We do not show the finite-size extrapolation, which proceeds similarly
to that in Figs.\ \ref{fig:spin_gap_L_U0_J20} and 
\ref{fig:charge_gap_L_U0_J20}, explicitly.)
The spin gap opens slowly at small $J_2$, but with a form consistent
with a critical $J_2^{s}=0$ (see the inset in particular).
The charge gap decreases rapidly with $J_2$ at small $J_2$, reaching zero at 
$J_2^{c(1)} \approx 1 = J_1$, but then 
opens again at $J_2^{c(2)} \approx 2$. 
At weak coupling, this behavior of both gaps is consistent with
the predictions of bosonization, 
but the reopening of the charge gap for larger $J_2$ is not contained
in the bosonization analysis.
However, such large values of $J_2$ are clearly outside its region of
validity.

Representative results for the finite-size scaling of the bond order
parameter $\langle B \rangle$ are present in 
Fig.\ \ref{fig:BOW_L_U0_J1_J2}. 
For small $J_2$, the scaling behavior is similar to that for  
$J_2=0$, yielding an exponent $\gamma$
% $\approx 0.44, \ldots , 0.71$. 
that varies between $0.44$ and $0.71$.
However, for large $J_2$, the data extrapolate almost linearly to finite
values.
This illustrates that the scaling form (\ref{eq:BOW_extrapol}), goes
over to a function that might be better fit by a polynomial in $1/L$,
as in Eq.\ (\ref{eq:Gap_extrapol}).
However, for consistency, we nevertheless always use 
 Eq.\ (\ref{eq:BOW_extrapol})
for the fitting and note that the case of a linear function of $1/L$
is encompassed by Eq.\ (\ref{eq:BOW_extrapol}) with $\gamma=1$.

\begin{figure}
\includegraphics[scale=0.35,clip=true]{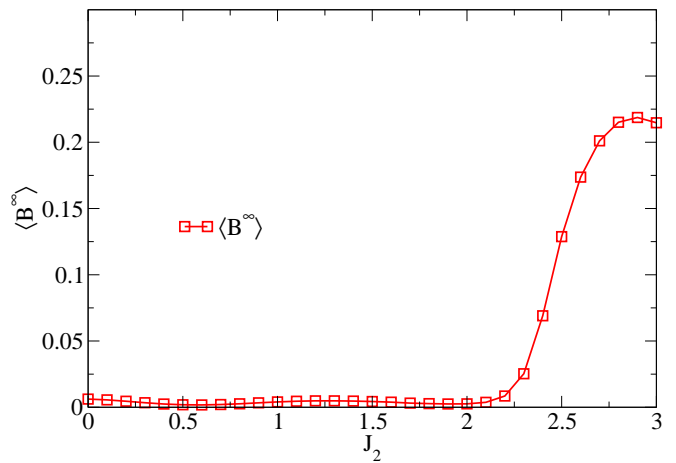}
\caption{(Color online) The $L = \infty$ extrapolated bond order parameter 
$\langle B^\infty \rangle$ as a
function of $J_2$ for $U=0$ and $J_1=1$. }
\label{fig:BOW_U0_J1_J2}
\end{figure}

The extrapolated results for $\langle B^\infty \rangle$, plotted as a 
function of $J_2$, are shown in Fig.\ \ref{fig:BOW_U0_J1_J2}. 
For $J_2 = 0$ to $J_2^{c(1)}$, $\langle B^\infty \rangle$ is very
small, even falling off from the small finite value at $J_2 = 0$,
which we have argued to come about due to numerical and extrapolation
errors.
Note that here, for $J_2 < J_2^{c(1)} \approx 1$,
the phase is characterized as bond order wave within bosonization.
While this seems to be a contradiction at first glance, 
note that the charge gap, Fig.~\ref{fig:gaps_U0_J1_J2}, 
falls off very rapidly from its small finite value at $J_2 = 0$,
whereas the spin gap opens very slowly due to its putative exponential
form.
In consequence, the value of $\langle B^\infty\rangle$ 
is very small. 
Our interpretation, then, is that the BOW order parameter is
finite, but numerically unresolvable in this region.  
For $J_2^{c(1)} < J_2 < J_2^{c(2)}$, the spin gap is
clearly non-vanishing, but $\langle B^\infty \rangle$ is numerically zero.
This behavior is consistent with the bosonization prediction
of a Luther-Emery phase. 
In other words, the vanishing charge gap indicates a phase in which
there is no BOW.
When $J_2 > J_2^{c(2)}$, coincident with the reopening of the charge
gap in Fig.~\ref{fig:gaps_U0_J1_J2}, the BOW phase reappears, this
time clearly marked by a finite bond order parameter as well as finite
spin and charge gaps.

\subsection{Results for nonzero $U$ and $J_2$}
\begin{figure}
\includegraphics[scale=0.35,clip=true]{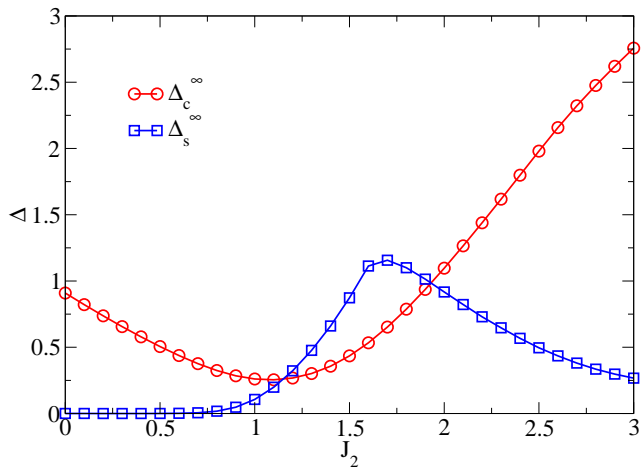}
\caption{(Color online) The $L = \infty$ extrapolated spin gap and charge gap as
functions of $J_2$ for $U = 2$, $J_1 = 1$. } 
\label{fig:gaps_U2_J1_J2}
\end{figure}

\begin{figure}
\includegraphics[scale=0.35,clip=true]{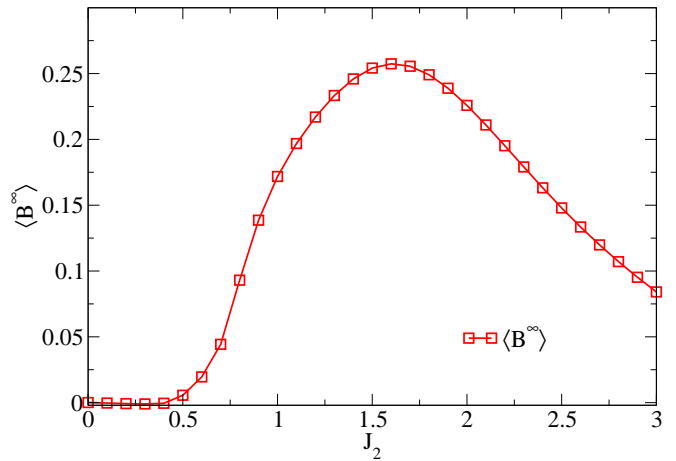}
\caption{(Color online) The $L = \infty$ extrapolated $\langle B \rangle$ as a
function of $J_2$ for $U = 2$, $J_1 = 1$. }
\label{fig:BOW_U2_J1_J2}
\end{figure}

We now study the effect of the frustration $J_2$ when the Coulomb
repulsion $U$ is finite.
Bosonization predicts that the SDW phase that is present
only along the $J_2=0$ line at $U=0$ becomes enlarged to a finite region at
finite $U$.
We explore the behavior as a function of $J_2$ 
for moderate values of $U$ and $J_1$, $U = 2$, and $J_1 = 1$.
Fig.\ \ref{fig:gaps_U2_J1_J2} shows
the spin and charge gaps, extrapolated to infinite systems size, as a
function of $J_2$.
As can be seen, the spin gap opens 
at a finite $J_2^s \approx 0.6$ and the charge gap, although at first
decreasing and reaching a minimum at $J_2\approx 1.1$, is always
finite. 
%The extrapolation behavior of the $\langle B \rangle$ here is also as 
%above Fig.\ref{fig:BOW_L_U2_J1_J2}. 
As can be seen in Fig.\ \ref{fig:BOW_U2_J1_J2}, the bond order
parameter $\langle B^\infty \rangle = 0$ when $J_2 < J_2^s$, and
opens rapidly to a large, finite value at $J_2 \approx 0.5$.
The behavior of all quantities is consistent
with a SDW phase for small $J_2$ and a BOW for large $J_2$.
Bosonization does predict a transition from a SDW phase to a BOW
phase at $J_2 = U/2$ (see Fig.\ \ref{fig:phases-RG}).
However, it also predicts a 
transition to a spin-gapless LE phase at larger $J_2$, which is not
found in the numerical calculations.
In our opinion, this is because the values of $U$, $J_1$ and $J_2$
here are large enough so that the regime of validity of bosonization
is exceeded.
Note that the critical value $J_2^s \approx 0.5$ is far from the
weak-coupling prediction of $J_2 = U/2 = 2$, but agrees fairly well with
the value expected from the frustrated Heisenberg chain, for which
$(J^{\text{Heis}}_2/J^{\text{Heis}}_1)_c \approx0.241$,
\cite{eggert_1996,okamoto_1992} 
if we take 
$J_1^{\text{Heis}}=J_1 + 4 t^2/U = 3$, the effective Heisenberg coupling
within strong coupling; this yields an estimate
$J_2^{c(\text{strong})} \approx 0.72$, in reasonable agreement with the
DMRG result.
  
\begin{figure}
\includegraphics[scale=0.35,clip=true]{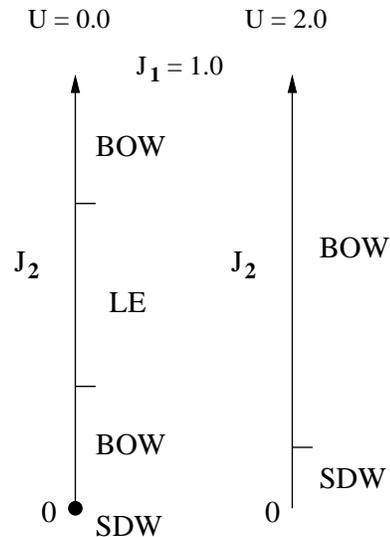}
\caption{A sketch of the ground-state phase diagram of the one-dimensional 
$t$-$U$-$J_1$-$J_2$ model at zero and finite $U$ obtained from analysis of
the DMRG calculations. }
\label{fig:DMRG_Phase}
\end{figure}

Fig.\ \ref{fig:DMRG_Phase} summarizes the phase diagrams as a function
of $J_2$ obtained from the DMRG calculations at zero and finite $U$.
For $U=0$, the SDW phase at $J_2=0$ becomes a BOW phase at arbitrarily
small, but weak $J_2$.
At intermediate $J_2$, a metallic, but spin-gapped Luther-Emery phase
occurs, and at large $J_2$ the system reenters the BOW phase.
At moderate, finite $U$, the SDW phase persists when $J_2$ is small
and finite, going over to a BOW at larger $J_2$.

\section{Discussion and conclusion}
\label{sec:discussion}
In this work, we have investigated the ground-state behavior of the
half-filled one-dimensional Hubbard model with antiferromagnetic
nearest-neighbor 
and next-nearest-neighbor Heisenberg interactions.
Our field-theoretical analysis for weak couplings indicates
that the ground state has a finite gap for either charge excitations
(spin-density-wave phase, SDW) or spin excitations (Luther-Emery phase, LE) 
or both (bond-order-wave phase, BOW).
Our extensive numerical DMRG investigations
agree very well with the field-theoretical predictions for small interactions.
The only exception is the lack of numerical evidence
for a finite bond-order parameter 
in the region $U=0$, $J_1=1$ and $0<J_2<J_1$. Here the system sizes
are large enough to resolve finite spin and charge gaps but
they are still too small to detect the very small bond order parameter.

For larger interactions, e.g., $U=2$, the DMRG finds a strong-coupling
bond-order-wave phase which eludes the field-theoretical description.
Instead, its existence and its properties can be inferred from
a strong-coupling expansion of the model where it is seen
that the strong-coupling BOW phase results from the
frustration of the nearest-neighbor and next-nearest-neighbor 
Heisenberg couplings.
Therefore, the metallic Luther-Emery phase is limited to
a narrow weak-coupling region in the phase space where it would be 
very difficult to justify the strengths of the coupling parameters 
from microscopic considerations.
For moderate interactions, an echo of the weak-coupling
Luther-Emery phase can be seen
in the behavior of the charge gap as a function of $J_2$, which
displays a minimum at some $J_2\gtrsim J_1$.

The nearest-neighbor Heisenberg coupling $J_1$ is {\em not\/} 
a frustrating interaction for the half-filled Hubbard model
because the ground state of the $t$-$U$-$J_1$ model is a spin-density wave
for all $J_1\geq 0$. 
In order to arrive at this conclusion
in the field-theoretical analysis,
the fact that bosonic phase fields 
are locked to their mean-field values when excitations are gapped, so that 
seemingly irrelevant operators become marginal operators, must be taken
into account. 
In numerical calculations one needs to study rather large system sizes
in order to extrapolate to a vanishing spin gap and bond-order parameter
in the thermodynamic limit.
The next-nearest-neighbor Heisenberg interaction $J_2$, in contrast,
truly frustrates the Hubbard model, opening the way to 
Luther-Emery and bond-charge-ordered phases for $J_2>0$. 
As expected
from our experience with the frustrated Heisenberg model,
the SDW phase is stable against weak frustration for $U>0$, i.e.,
a finite $J_2$ is required to open the spin gap. 

In conclusion, our study demonstrates both analytically and numerically
that a nearest-neighbor Heisenberg exchange
interaction added to the half-filled Hubbard model does not lead to
frustration or to new phases in the ground-state phase diagram,
whereas a frustrating next-nearest-neighbor exchange does.

\acknowledgments{This work was partly
  supported by the Hungarian Research Fund (OTKA) Grant No. K-68340, and by
the DFG-OTKA International Research Training Group 790 
{\sl Electron-Electron Interactions in Solids}.}

\end{document}